\documentclass{aa}
\usepackage{graphics}

\newcommand{\msun}{\,$M_{\odot}$}

\begin{document}

\thesaurus{04         
           (04.19.1); 
           08         
           (08.12.2;  
            08.16.5); 
           10         
            (10.15.2)}

\title{CCD-based observations of PG\,0856+121 and a theoretical analysis 
       of its oscillation modes\thanks{Based on observations made 
         with the IAC80 Telescope operated on the island of Tenerife by 
         the Instituto de Astrof\'\i sica de Canarias in the Spanish 
         Observatorio del Teide.} }

\subtitle{}

\author{A. Ulla\inst{1}, M. R. Zapatero Osorio\inst{2,3}, 
        F. P\'erez Hern\'andez\inst{2,4}, J. MacDonald\inst{5}}

\offprints{A. Ulla}
\mail {ulla@uvigo.es}

\institute{Universidade de Vigo. Departamento de F\'\i sica Aplicada. 
           \'Area de F\'\i sica da Terra, Astronom\'\i a e Astrof\'\i sica. 
           Facultade de Ciencias, Campus Lagoas-Marcosende, E-36200 Vigo, 
           Spain
    \and   Instituto de Astrof\'\i sica de Canarias. c/ V\'\i a L\'actea 
           s/n, E-38200 La Laguna, Tenerife, Spain
    \and   CALTECH, MS 150-21, Pasadena, CA 91125, USA
    \and   Departamento de Astrof\'\i sica, Universidad de La Laguna, 
             Tenerife, Spain
    \and   Department of Physics and Astronomy, University of Delaware, 
           Newark, DE 19716, USA
          }

\date{Received ; accepted }

\authorrunning{Ulla et al.}

\titlerunning{PG\,0856+121: Observations and a theoretical analysis.} 

\maketitle

\begin{abstract}
$BVRI$ CCD-based and near-IR ($J$) imaging, together with unfiltered photometry of the hot subdwarf B star PG\,0856+121 are 
reported. Two close, faint, red, point-like sources are resolved. 
They account for the previously reported IR excess observed in this 
hot subdwarf. 
In addition, the new unfiltered differential photometry of PG\,0856+121 confirms its previously reported pulsational nature. A comparison with the oscillation modes
of stellar models suggests the possible presence of g modes. 
\end{abstract}

\section{Introduction}

Hot B-type subdwarfs (sdBs) are H-rich blue subluminous objects with 
temperatures 
not exceeding about 35000\,K (Greenstein \& Sargent 1974; Heber 1986). 
They have a canonical mass of 0.55\msun, with thin H-rich envelopes of less 
than 0.02\msun, and a distribution in log\,$g$ around 5.25--6.5 (Ulla \& 
Thejll 1998, hereafter referred to as  UT98). These objects are proposed as 
likely progenitors of white dwarfs and descendants of blue borizontal branch 
stars or asymptotic giant branch (AGB) stars (Saffer et al. 1998). They are 
also proposed to be responsible for the UV upturn flux observed in 
early-type galaxies (Bica et al. 1996). Among the various theories for the 
origin and final fate of the sdBs, close binary evolution has been suggested 
as one of the likely channels. These investigations are relevant to the 
formation of Type Ia supernovae by merging of double-degenerate pairs, in 
which one or both members could be descendants of hot subdwarfs 
(Saffer et al. 1998). Enough evidence has been accumulated to date in favor 
of a binary nature for 
at least 40\% of the field hot B subdwarf stars (e.g. Allard et al. 1994; 
Jeffery \& Pollacco 1998; UT98), with the detected companions ranging 
broadly in spectral type and physical parameters.

It is therefore very important to continue to seek information on the 
current binary nature of hot subdwarfs. In this regard and based on $JHK$ 
photometry, Thejll et al. (1995, hereafter referred to as TUM95) and UT98 have compiled a list of 
suitable candidates. They also suggested some particular targets for 
further investigation, despite large error bars associated with the 
observations (see UT98 for details). A way to pursue a more detailed study 
of the binary nature of such objects is to obtain filtered CCD imaging 
to search for close red components whose IR emission could have contributed 
to TUM95 and UT98 measurements. With that aim we have started such a 
program and present here results for the sdB star PG\,0856+121. 
Table 1 summarizes relevant information about this object published to date. 

PG\,0856+121 was suspected to be a possible pulsating sdB candidate by Koen et al. (1997, 1998a), based on the similarity of its physical properties to those of known sdB pulsating --or EC14026-- stars (Kilkenny et al. 1997). This suspicion was confirmed by Piccioni et al. (2000), who found periodic light variations with frequencies of 2.3\,mHz and 3.2\,mHz at a reasonable confidence level. We present our CCD and near-IR observations in Section 2. Section 3 provides a brief description of the models used and analysis performed to investigate the oscillatory nature of PG\,0856+121. Our conclusions are presented in Section 4.

\section{Observations and related results}

CCD-based images (1024\,$\times$\,1024\,pixels) in the Johnson $BVRI$ filters were obtained for PG\,0856+121 using the Thomson camera mounted on the Cassegrain focus of the 0.8-m IAC80 telescope (Teide Observatory) on April 13, 1998. The pixel size of the detector is 0.4325\arcsec. The night was photometric with an average seeing value around 2\arcsec. Raw frames were processed using standard techniques within the IRAF\footnote{IRAF is distributed by National Optical Astronomy Observatories, which is operated by the Association of Universities for Research in Astronomy, Inc., under contract with the National Science Foundation.} (Image Reduction and Analysis Facility) environment, which included bias subtraction, flat-fielding and correction for bad pixels by interpolation with values from the nearest-neighbour pixels. Landolt (1992) standard stars were also observed at various air masses in order to convert instrumental magnitudes into absolute data. Average {\em rms} values for 
the photometric calibration of each filter are as follows: 0.022\,mag for $B$, 0.020\,mag for $V$ and $R$, and 0.018\,mag for $I$. Photometry for the 
target star PG\,0856+121 has been achieved via the 
point spread function (PSF) fitting method. The stellar PSF was 
determined for each colour frame using at least three ``isolated'',
bright point-like sources (not our target) appearing in the IAC80
images, and it was later applied to PG\,0856+121 providing the
following colours and magnitudes in the Cousins photometric system:
$V$\,=\,13.559, $B-V$\,=\,$-0.311$, $R$\,=\,13.667, $R-I$\,=\,$-0.138$.

\begin{figure}
\input epsf
\centering
\leavevmode
\epsfxsize=1.0
\columnwidth
\epsfbox{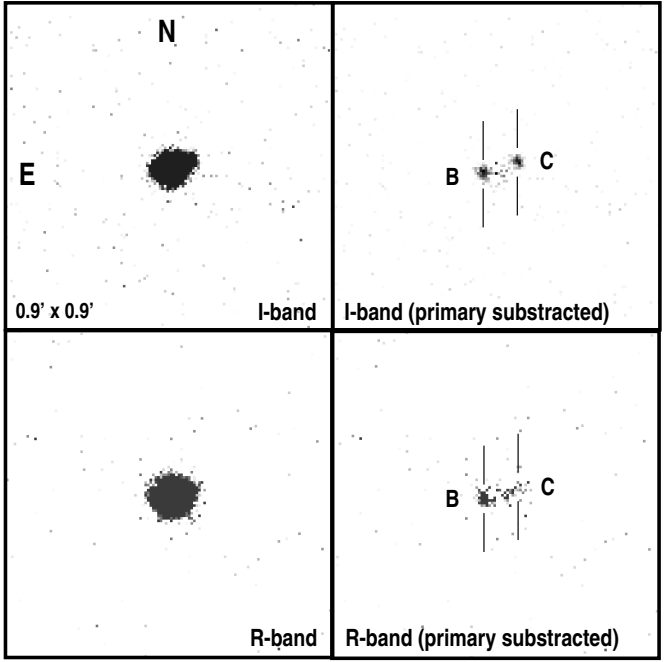}
\caption{R- and I-band CCD images of PG0856+121 in which the locations of 
the "B" and "C" point-like sources are indicated.} \label{f2}
\end{figure}

Looking at the deconvolved $R$ and $I$ images of PG\,0856+121 it became evident that two faint red objects are present very close to the target; one object, named ``B'' ($I\,\le\,17.1$, $R\,\le\,17.4$), located at $\le$\,2.4\arcsec~eastward of our star, and another one, named ``C'' ($I\,=\,17.2\,\pm\,0.05$, $R-I\,=\,1.2-1.6$), located at about 3.5\arcsec~northwest (P.A.\,=\,296$^{\circ}$) from it. If object ``C'' is a Main Sequence star of solar metallicity, a spectral type of M2--M4 would be inferred for it. Figure~1 shows the $R$ and $I$ images of PG\,0856+121 in which the locations of the ``B'' and ``C'' point-like sources are indicated. Both sources are quite well detected when subtracting the average PSF from the central target. The typical PSF has FWHM values of 2.2\arcsec~ and 1.9\arcsec~in the $R$ and $I$ frames, respectively. In view of this discovery, we now interpret the suggestion of UT98 for a binary nature of PG\,0856+121 in a different way: as they employed a 15\arcsec~aperture for their $JHK$ observations of the target, it is now clear that the two nearby red objects contributed significantly to their measurements. In particular, based on UT98 $JHK$ values for PG\,0856+121 (see Table 1) and under the assumption of ``C'' being a typical M-type field star, a contribution of 15\%~in the $J$ band and of about 40\%~in the $K$ band can be estimated. To test whether either of the two nearby objects is gravitationally linked to PG\,0856+121 by taking long-term radial velocity data sets is beyond the scope of the present work. In any case, we might be dealing with a detached system whose long period and orbital parameters should be tested against close binary evolution theories proposed for the hot subdwarfs (see, e.g., Iben \& Tutukov 1986a,b or Iben 1990). 

\begin{figure}
\input epsf
\centering
\leavevmode
\epsfxsize=1.0
\columnwidth
\epsfbox{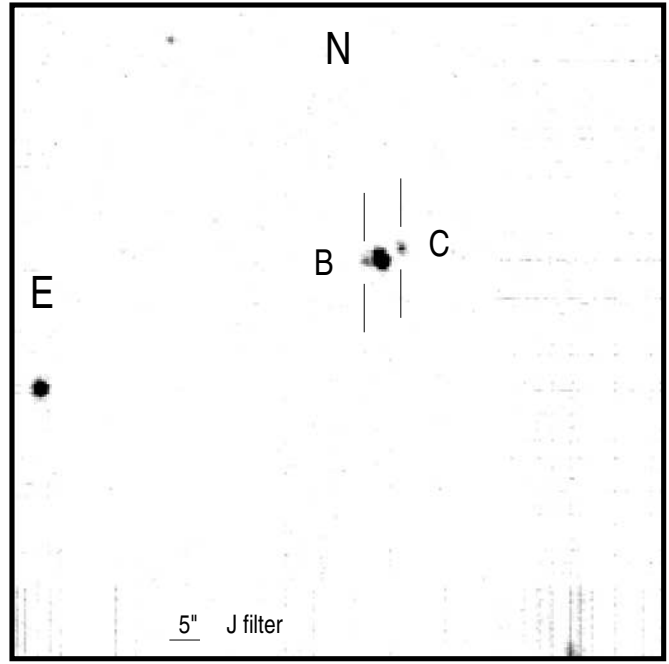}
\caption{$J$-band image of PG\,0856+121 (central, bright star) confirming
   the presence of two nearby, faint sources labeled as "B" and "C"
   (see text).} \label{f22}
\end{figure}

On November 25, 2000, with the goal of proving objects ``B'' and ``C'',
$J$ images of PG\,0856+121 were obtained with the near-infrared
camera (Hg\,Cd\,Te detector, 256\,$\times$\,256) mounted on the
Cassegrain focus of the 1.5-m Carlos S\'anchez Telescope (Teide
Observatory). We performed the observations through the
``narrow-optics'' of the instrument which provides a pixel projection
of 0.4\arcsec~onto the sky. The atmospheric seeing conditions during
the night of the observations were fairly stable around
1.3\arcsec. The total integration time was 600\,s, the final $J$ image
(Fig.~2) being the co-addition of five dithered exposures of 120\,s
each. Objects ``B'' and ``C'' around PG\,0856+121 are clearly   
resolved in this image, and thus proved to be real. The
astrometric measurements carried out on this frame confirm those of
the IAC80 observations for object ``C'' and provide for object
``B'' a separation of 2.3\arcsec~at a position angle of 98$^{\circ}$ from
PG\,0856+121. Instrumental photometry has been performed using a
similar procedure as described above and calibrated into real
magnitudes with the observations of the standard star AS19-1 (Hunt et
al. 1998), which were taken with the same instrumental configuration
just after our target. The combined $J$ magnitude of the three sources
(PG\,0856+121 and objects ``B'' and ``C'') is 13.89\,$\pm$\,0.10\,mag;
this value is in agreement within the error bars of the near-infrared 
photometry given in UT98 for PG\,0856+121. Nevertheless, we have
derived the $J$ photometry for each star: 14.16\,mag (78.3\%~is the
contribution to the combined flux) for PG\,0856+121, 16.59\,mag
(8.3\%) for object ``B'', and 16.08\,mag (13.3\%) for object ``C''.
This latter object is the reddest of the three, and its contribution
to the light in the near-infrared is indeed significant. Our $J$ data
confirms object ``C'' as an early- to mid-M type dwarf, and would
place it at a distance of 750\,$\pm$\,350\,pc (adopting a main
sequence calibration). This is only marginally consistent with the
estimated distance of PG\,0856+121. The $I-J$ colour of the hot
subdwarf PG\,0856+121 is now $-0.36$\,mag, which compares well to
typical colours of other B type hot subdwarfs.

Unfiltered CCD photometry of PG\,0856+121 was also performed on February 
27, 2000, using the same camera and telescope as for our previous 
observations (on April 13, 1998). The target was monitored every 37\,sec (25\,sec integration time plus overheads) during 2.4 hours in the airmass interval 1.04--1.20. The CCD detector was windowed so that two comparison stars of similar brightness in the field were observed simultaneously. We performed differential photometry of our target with an accuracy of the order of 0.005\,mag. In brief, the procedure was as follows: apertures for PG\,0856+121 and the two reference stars were defined as a function of the average FWHM of the frame, and the sky intensity was set as an outer ring of width 1.5\,pixel. We compared the reference stars against each other and found them to be constant at the level of our 1\,$\sigma$ photometric error bars. 

\begin{figure}
\input epsf
\centering  
\leavevmode
\epsfxsize=1.0
\columnwidth
\epsfbox{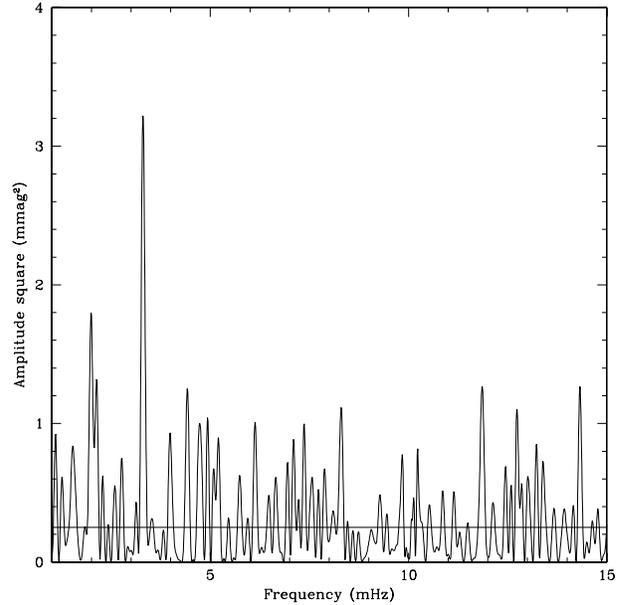}
\caption{Frequency spectrum for PG\,0856+121}
\label{f0} 
\end{figure}

Spectral analysis of the February 27, 2000, data was achieved using the ISWF method over the frequency range 1\,mHz to 15\,mHz. The frequency spectrum is shown in Fig.~\ref{f0} (amplitude square versus cyclic frequency).
The horizontal line is the square of the mean value of the amplitude in the
frequency range from 1\,mHz to 15\,mHz. The frequency resolution is 0.12\,mHz. 
We found a significant peak at 3.30\,mHz (303\,sec) with a signal-to-noise 
ratio of 3.6 in amplitude (roughly a 99\% confidence level). 
This frequency peak does not differ within uncertainties from that derived by 
Piccioni et al. (2000), {\em {i.e.}} 3.2 mHz. This confirms the oscillatory nature of  PG\,0856+121. 
Although with a smaller confidence level, peaks at 2.13\,mHz 
(469.5\,sec, signal-to-noise of 2.3 in amplitude), and 1.99\,mHz (502.5\,sec, 
signal-to-noise of 2.7 in amplitude) can be related to the one found by Piccioni
et al. (2000) at 2.3\,mHz with a similar signal-to-noise level.
However, given the confidence level of our own measurements and those of 
Piccioni et al. (2000), we caution that the reliability of the shorter frequencies should be confirmed by further observations.

\section{Pulsational model calculations}

Oscillations in hot sdB stars have now been well established by
observations (see O'Donoghue et al. 1998 and references therein).
Potentially, this allows analysis of the internal structure of sdB stars by
comparison of the observed frequencies with those corresponding to stellar
models with different physical assumptions. Some work has already been done in this direction (see, e.g., Charpinet et al., 1997; Billeres et al., 1998; or Ulla et al. 1999).

Here we compare the frequency peaks reported in the previous section and 
those found by Piccioni et al. (2000)
with theoretical frequencies based on stellar structure
models compatible with the surface parameters of PG0856+121. 
We have computed stellar structure models of different masses, suitable 
for the sdB star PG0856+121. The equation of state, opacity and nuclear reactions are briefly described in Jim\'enez \& MacDonald (1996). An 
additional change is the use of OPAL95 opacity tables (Iglesias \& Rogers,
1996). These models have helium cores and thin H-rich envelopes. A summary of the models here considered is given in Table 2.
To produce surface abundances similar to those in PG0856+121, we have
included gravitational settling and element diffusion (Iben and MacDonald
1985) in models 4, 6 and 8.
The envelope compositions for models 1 and 2 are X=0.71, Y=0.29, Z=0.0001
and for models 3, 5 and 7 X=0.60, Y=0.38, Z=0.02. For the models with
diffusion the initial envelope composition is also X=0.60, Y=0.38, Z=0.02.Gravitational settling causes helium 
and heavy elements to quickly 
sink below the photosphere. The outer layers are then pure hydrogen. 
The fact that n(He)/n(H) = 0.01 in PG0856+121 and other sdBs can be 
explained by the presence of a wind that counters the effects of 
gravitational settling. The wind mass loss rates required to do this 
are quite small (10$^{-15}$ - 10$^{-14}$ M$_\odot$/year) and completely undetectable up to date with ordinary techniques and instrumentation.

The models eigenfrequencies were computed in the
adiabatic approximation, using the code developed by
Christensen-Dalsgaard (see Christensen-Dalsgaard \& Berthomieu, 1991).
For models with similar internal structure, as in the present case, the
dynamical time scale $t_{\rm dyn}=(R^3/GM)^{1/2}$ dominates the
variation of the oscillation frequencies. Thus it is convenient to
compare our results in terms of dimensionless frequencies
$\sigma$, defined by

\begin{equation}
\sigma \equiv \left(\frac{R^3}{GM}\right)^{1/2} \omega \; ,
\end{equation}

\noindent
where $\omega$ is the angular oscillation frequency and the other symbols have their conventional meanings.

\begin{figure}
\input epsf
\centering  
\leavevmode
\epsfxsize=1.0
\columnwidth
\epsfbox{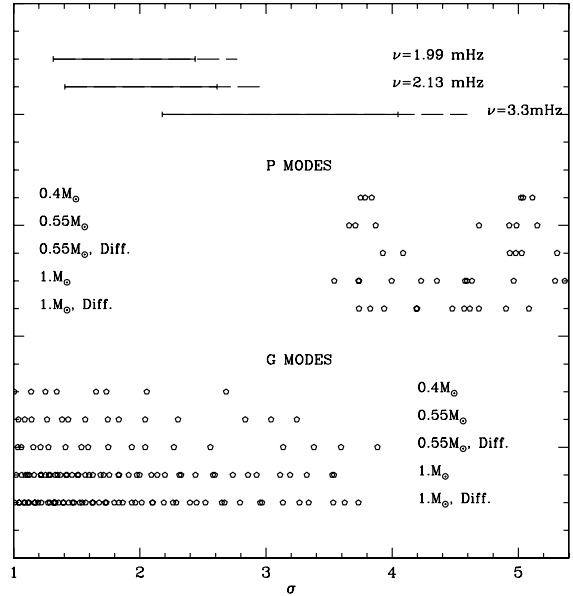}
\caption{Dimensionless frequencies for PG0856+121. Only 5 models, of those in Table 2, are shown for clarity.}
\label{f1} 
\end{figure}

In Fig.~\ref{f1} we show the theoretical dimensionless frequencies,
computed
by using equation (1). Only modes with $\ell\leq 2$ are considered since
modes of higher degree can hardly be observed for point-like stars.
For the model with $0.4M_{\odot}$, the 6 p-mode frequencies shown in 
Fig.~\ref{f1} correspond to the fundamental and first overtone of 
the $\ell =0,1,2$
degrees. For other models, the p-mode spectrum is more complex due to
the presence of g-like modes. 
As it can be seen in the figure, the dimensionless frequencies decrease with mass and increase when diffusion is considered.

The dimensionless frequencies corresponding to the observational periods are
usually estimated by expressing $\sigma$ in terms of $T_{\rm eff}$,
$\log g$ and the luminosity $L$ of the stars.
However, for this kind of star it seems better to use some estimate of the
mass rather than of $L$. In fact, the distance quoted in Table 1 was obtained
by assuming the canonical mass for sdB stars.
The relations are

\begin{equation}
\sigma= \left( \frac{L}{4\pi\sigma_{\rm SB}}\right)^{1/4}
\frac{\omega}{g^{1/2}T_{\rm eff}} = \left(\frac{GM}{g^3}\right)^{1/4} \omega
\end{equation}

\noindent
where $\sigma_{\rm SB}$ is the Stefan-Boltzmann constant.

In particular, for PG0856+121,
we have used the value $\log g = 5.73 \pm 0.15$ (Saffer et al. 1994).
For the mass we use the canonical value
$0.5\pm 0.1 M_{\odot}$ (Saffer et al. 1994). 
In addition, the observational frequencies have associated errors, but they
are negligible when compared to those of $\log g$ and $M$.
The resulting dimensionless frequencies for the observed frequency peaks
reported here are shown at the top of Fig.~\ref{f1} as horizontal 
continuous lines. Use of the results of Piccioni et al. (2000) instead 
of those shown in the figure, does not change the following discussion.
The horizontal dashed lines in Fig.~\ref{f1} were obtained by assuming the
same uncertainty in $\log g$ as quoted above but for a mass range of $0.4M_{\odot} < M < 1 
M_{\odot}$. This allows us to explore the possibility that PG0856+121 
has a mass substantially larger than the canonical one.

In the previous analysis the frequency splittings caused by rotation have
been neglected. Although we do not know the rotational velocity of this
particular star, we shall consider the value 90 km/s as an upper limit. This
follows from the work by Saffer et al. (1994; and a private communication), who measured this quantity for about 50 sdB stars and none was found to be rotating faster than about 90 km/s. Then, by
using the stellar radii given in Table 2,  it can be seen that the value of the
rotational frequency is, at most, $5\%$ that of the observational frequencies
and, hence, first order corrections for the frequency splitting will be enough for
our purposes. By using the values of $M$ and $R$ in Table 2, a rotational
frequency splitting $\beta_{nl} m \sigma_{\rm rot}$ (here $m$ is the azimuthal 
order, $\sigma_{\rm rot}$ the dimensionless rotational angular frequency and 
the parameter $0\leq\beta_{nl}\leq 1$) smaller than
0.25 is found for modes with $\ell \leq 2$. Considering this additional
uncertainty in Fig.~\ref{f1} also does not change the conclusions given below. 
                                                         
From Fig.~\ref{f1} it follows that the peak at 3.3 mHz can be either a g mode 
or a p mode. Since in other sdB stars only p modes are detected (see e.g. 
Billeres et al. (1998), Koen et al. 1998b) in agreement with the theoretical 
expectations of Fontaine et al. (1998) for the EC14026 stars, the latter 
possibility can be considered with preference. In this case, and 
assuming the canonical mass, the peak at 3.3 mHz would be a fundamental 
p-mode with degree $\ell=0,1$, or $2$.
On the other hand, if any of the peaks at 1.9 and 2.1 mHz are real and the photometric value
of $\log g$ is correctly determined, from
Figure~\ref{f1} it follows that these peaks must be g modes of low
order. 
It is important to note that this conclusion does not depend on the details
of the model structure, but only on the stellar parameters.
The basic reason is that the dimensionless frequencies $\sigma$ of the p modes
are, in a first approximation, independent of such details.

\section{Summary and Conclusions}

Optical ($BVRI$) and near-IR ($J$) imaging of the field nearby the sdB 
star PG0856+121 has revealed the presence of two faint red objects very 
close ($\le$\,4\arcsec) to the target. In view of this discovery, 
contamination by them in the 
near-IR bands is here proposed as the most likely interpretation for the
previously reported $JHK$ values for the object (UT98). Our photometric 
data show that the optical and near-IR colours of PG\,0856+121 are 
consistent with those of other single hot subdwarfs. Whether either of 
the two red objects is gravitationally linked to PG0856+121 has not been 
investigated further but it is now brought to the attention of 
potentially interested researchers. If a binary nature could be 
established for the target, a refinement in the determination of its 
physical properties together with those of its companion would be 
obtainable, in the ways abundantly documented in the literature already. 
It is worth noting that if the given distance (Table 1) of 990 pc
to the target is correct, then the reddest "C" companion, for which an 
early- to mid-M type dwarf has been determined given its IR colours, 
would be placed at a distance of 750\,$\pm$\,350\, marginally 
consistent with the estimated distance of PG\,0856+121 above. This makes 
it very difficult to measure radial velocities by the usual spectroscopic
techniques. On the other hand, the presence of an even 
closer (and therefore unresolved) companion to PG0856+121 could still be
revealed through radial velocity measurements. 
As a further suggestion, checking of the eventual binary nature of
PG0856+121 towards either the "C" or "B" objects, could also be possible on an
approximate time-scale of 10 years by means of its proper motion values,
as provided by de Boer et al. (1997) and Colin et al. (1994).

Recently, PG0856+121 has been reported to display a pulsating nature by 
Piccioni et al. (2000). The new differential photometry of the target 
presented here mostly confirm the peaks, at 2.3 and 3.2 mHz, detected by 
these authors. In both works, the largest frequency peak is found at 
$3.3\pm 0.1$\,mHz and, also in both works, peaks around 2.0-2.3 mHz are 
found, although with a lower confidence level. We have compared these 
frequency peaks with those of stellar models compatible with the physical 
properties of the target star. Our results indicate that the peak at 
$3.3$\,mHz is a p or g mode with a low radial order; in particular, if 
the p-mode character is assumed and the canonical mass is considered, it 
would be the fundamental mode with degree $\ell=$0,1 or 2.

However, the other frequency peaks are in the g-mode range of our models.  
Since other pulsating sdB stars seem to have only p modes, in agreement 
with earlier theoretical computations, alternative explanations for our low 
frequency peaks need to be considered. 
First, the S/N ratio for these peaks is rather small and, hence, require 
further confirmation.  Also we note that the results are based on the 
stellar parameters provided by Saffer et al. (1994) --mainly the value of 
$\log g$--. To search for systematic shifts in these parameters would 
demand not only spectra of higher S/N ratio but also to test effects such 
as the importance of considering NLTE atmospheric models for the sdB 
stars. In view of the importance of these considerations for the 
particular case of PG0856+121, we propose it as a candidate for future 
improved spectroscopic studies in an attempt to further constrain its oscillatory properties.
On the other hand, we find it unlikely that this result arises from 
errors in the stellar models analysis performed; in particular, as 
indicated in Section 3, even if the models considered were unsuitable for 
the target, the p-mode range of frequencies has a small dependence on 
the models' details, and a broad mass range (up to $1 M_{\odot}$) has 
been tested still yielding low frequency peaks in the g-mode range.

\section*{Acknowledgments}

We are grateful to V.\,J.\,S. B\'ejar for his assistance in obtaining the 
$J$ images. 
Ana Ulla acknowledges support from the Spanish MEC DGESIC under contract
PB97-1435-C02-02 and from the Spanish MCT DGI under contract AYA2000-1691. 
This research has made use of the SIMBAD 
database, operated at CDS, Strasbourg, France and of the NASA Astrophysics 
Data System (ADS).

\begin{table*}
\caption{Summary of properties of PG0856+121 (references provided).}
\begin{center}
\begin{tabular}{lcccc}
\hline
{\bf Names:}& PG0856+121& (1)& WD0856+121& (2) \\
{\bf Sp.type:}& sdB& (1; 3)& non variable& (4) \\
{\bf RA}(1950)= & 08 56 18.8  & {\bf Dec} =& +12 08 06   & (1) \\
         (2000) & 08 59 02.723&    & +11 56 24.73& (5) \\
{\bf LII} =          & 216.56      & {\bf BII} =& 33.67       & (3) \\
{\bf l} =         & 216.49      & {\bf b } = & 33.68       & (5) \\
{\bf distance:}     & (pc)        & 990& $\pm$ 370   & (5, 6)\\
{\bf heigh:}        & (pc)        & 550& $\pm$ 210   & (3) \\
{\bf rad.vel.:}& (km/s)      & +85&             & (5, 6)\\
                   &              & +97& $\pm$ 10.2  & (4) \\
{\bf prp.mot.:}& $\mu_\alpha$cos$\alpha$& $\mu_\delta$& (mas/yr)& \\
                   &             -19.4      &    -19.8    &     &  (5, 6)\\
\multicolumn{5}{l}{{\bf orbital components, velocity and parameters:} (5)}\\
  X = -9.16  & Y = -0.49  & Z = 0.55  & (kpc)  & \\
    U = -74  & V = 116    & W = -46   & (km/s) & \\
 It (kpc km/s)=     & -1099 & ecc = 0.48& nze = 0.14 & \\
\multicolumn{5}{l}{{\bf magnitudes:}}\\
      {\bf B } &    {\bf V }&  {\bf R} &  {\bf I} & \\
     13.248    &   13.559   &  13.667  & 13.805 &  \\
      $\pm$0.022& $\pm$0.020&$\pm$0.020& $\pm$0.018 & (7) \\
       {\bf B } &  {\bf P$_{mag}$}& {\bf U-B}  & {\bf B-V}  &        \\
      13.28  &  13.03    & -1.03& -0.19& (1) \\
       {\bf v} & {\bf u-v} & {\bf g-v} & {\bf g-r}&         \\
        13.52& -0.11     & -0.15& -0.52& (1)\\
          {\bf y} &  {\bf b-y} & {\bf m1}& {\bf c1}&      \\
        13.47&  -0.094   & +0.106& -0.004&   (8)\\
        13.50&  -0.116   & +0.113& +0.021&   (8)\\
    13.473   & -0.095    & -0.094& +0.035&       \\
   $\pm$0.012& $\pm$0.004&$\pm$0.005& $\pm$0.007&  (9)\\
   13.495    & -0.116    & +0.113   & {\bf u-b} &   \\
   $\pm$0.030& $\pm$0.032& $\pm$0.036&  .015$\pm$.021 & (10)\\
         {\bf J} & {\bf H }&  {\bf K }&  {\bf E(B-V)} &   \\
       13.42 & 13.59     & 13.84    & $\le$ 0.025& \\
    $\pm$0.44& $\pm$0.22 & $\pm$0.55&            &  (11)\\
 {\bf f$_{6700}$/f$_{6050}$} = & 0.651& {\bf f$_{7050}$/f$_{6700}$} = & 0.840&
(12)\\ 
{\bf T$_{eff}$:} (K)& 22000& 23800& 33000&  (3)\\
                    & 26400&      &      &  (13)\\
{\bf log($g$):}     & 5.1& (9)& 5.73 ({\bf Y}=0.001)& (13)\\
\multicolumn{5}{l}{{\bf optical sp.:} (9); {\bf finder chart:} (1)}\\
\hline
\multicolumn{5}{l}{1: Green et al. (1986); 2: McCook \& Sion (1987); 3: Moehler
et al. (1990a);}\\
\multicolumn{5}{l}{4: Saffer et al. (1998); 5: de Boer at al. (1997); 6
: Colin et al. (1994);}\\
\multicolumn{5}{l}{7: this work; 8: Kilkenny et al. (1988); 9: Moehler et al. (1990b);}\\
\multicolumn{5}{l}{10: Wesemael et al. (1992); 11: UT98; 12: Jeffery \& Pollacco (1998);}\\
\multicolumn{5}{l}{13: Saffer et al. (1994).} \\
\hline
\end{tabular}
\end{center}
\end{table*}                          

\begin{table}
\caption{Properties of sdB models 1 through 8, suitable for PG0856+121.} 
\label{tbl-1}
\begin{center}
\scriptsize
\begin{tabular}{cccccccc}\\\hline
Model & Mass & log$g$& Envlp. Mass&  Radius & Central $\rho$ & $T_{eff}$& Luminosity\\
nr. & \msun&       & \msun      & x10$^{9}$ cm  & x10$^{4}$ gr/cm$^3$  & x10$^{4}$ K & L$_\odot$
     \\\hline
1& 0.3999 & 5.843 & 3.18 10$^{-3}$ & 8.724&  3.329 & 2.641 & 6.891\\
2& 0.4517 & 5.644 & 6.46 10$^{-3}$ & 11.69&  2.550 & 2.642 & 12.42 \\
3& 0.5563 & 5.401 & 6.1  10$^{-3}$  & 17.12&  1.699 & 2.651 & 26.94\\
4& 0.5563 & 5.340 & 5.0  10$^{-3}$  & 18.37&  1.699 & 2.560 & 26.97 \\
5& 0.7941 & 4.862 & 2.43 10$^{-2}$ & 38.05&  0.941 & 2.696 & 142.4 \\
6& 0.7941 & 4.821 & 2.07 10$^{-2}$ & 39.88&  0.941 & 2.644 & 144.7 \\
7& 1.0608 & 4.390 & 5.06 10$^{-2}$ & 75.75&  0.611 & 2.629 & 510.6 \\
8& 1.0608 & 4.312 & 4.53 10$^{-2}$ & 82.83&  0.612 & 2.541 & 532.7 \\
\hline
\end{tabular}
\end{center}
\end{table}   

{}
\end{document}